\documentclass[aps,prl,floatfix,twocolumn,footinbib,amsmath,amssymb]{revtex4}
\usepackage{amsmath,amsthm,amssymb}
\usepackage{latexsym,psfrag}
\usepackage[normalem]{ulem}
\usepackage{graphicx,color}

\begin{document}

\vspace*{-8mm}
\hbox to 1\textwidth{\hfill OUTP-13-24P}
\vspace{3mm}
\title{Enhanced non-perturbative effects through the collinear anomaly}

\author{Thomas Becher\,$^a$ and Guido Bell\,$^b$ }

\affiliation{${}^a$\,Albert Einstein Center for Fundamental Physics, Institute for Theoretical Physics,\\
University of Bern, Sidlerstrasse 5, 3012 Bern, Switzerland\\
${}^b$\,Rudolf Peierls Centre for Theoretical Physics, University of Oxford, 1 Keble Road, Oxford OX1 3NP, United Kingdom}
\begin{abstract}
We show that non-perturbative effects are logarithmically enhanced for transverse-momentum-dependent observables such as  $q_T$-spectra of electroweak bosons in hadronic collisions and jet broadening at $e^+e^-$ colliders. This enhancement arises from the collinear anomaly, a mechanism characteristic for transverse observables, which induces logarithmic dependence on the hard scale in the product of the soft and collinear matrix elements. 
Our analysis is based on an operator product expansion and provides, for the first time, a systematic, model-independent way to study non-perturbative effects for this class of observables. For the case of jet broadening, we relate the leading correction to the non-perturbative shift of the thrust distribution. 
\end{abstract}
\maketitle

\section{Introduction}

For precision studies at colliders, such as the extraction of the strong coupling constant $\alpha_s$ from measurements of event shapes, it is important to control non-perturbative effects. While power suppressed, they can be relevant because the suppression scale is often not the center-of-mass scale $Q$ of the collision, but the smallest perturbative scale in the problem. For thrust $T=1-\tau$, for example, non-perturbative effects are only suppressed by the soft scale $\tau Q$, which is significantly smaller than $Q$ in the two-jet region where the extraction of $\alpha_s$ is performed. In the region $\Lambda_{\rm QCD} \ll \tau Q\ll Q$, 
the leading non-perturbative effect is a shift \cite{Korchemsky:1994is,Dokshitzer:1995qm}
\begin{equation}\label{thrust:shift}
\frac{d\sigma}{d\tau}(\tau) =  \frac{d\sigma^{\rm pert}}{d\tau}\left(\tau- c_0 \frac{\cal{A}}{Q}\right)+ \dots\,,
\end{equation}
where $c_0=2$ and $\cal A$ is a non-perturbative matrix element defined below. Lee and Sterman were able to show that the same matrix element 
$\cal A$ also provides the leading non-perturbative effect in other event-shape variables \cite{Lee:2006nr}. They considered the angularities \cite{Berger:2003iw}
\begin{equation}\label{eq:rapdef}
\tau_a =\frac{1}{Q} \sum_i e^{-(1-a) |\eta_i|} |p_{i}^\perp|\,,
\end{equation}
where the rapidity $\eta$ and transverse momentum are defined relative to the thrust axis, and found that the non-perturbative shift for these variables is the same as in the case of thrust, except for the prefactor, which equals $c_a =2/(1-a)$. This relation was obtained earlier from a dispersive model for power corrections \cite{Berger:2003pk}, but in \cite{Lee:2006nr} it was derived in QCD. For $a=0$ the angularity is equal to the thrust $\tau$, and for $a=1$ it is twice the total broadening $B_T=b_T/Q$. However, the analysis of \cite{Lee:2006nr} is only valid for $a<1$. In the limit $a\to1$, the factorization theorem \cite{Berger:2003iw} on which their analysis is based is no longer valid and the shift $c_a$ diverges. 

The factorization of event shapes in the two-jet limit can be derived using Soft-Collinear Effective Theory (SCET) \cite{Bauer:2000yr,Beneke:2002ph}. There has recently been a lot of progress in analyzing transverse-momentum-dependent observables in SCET. In particular, a factorization theorem for jet broadening was established in \cite{Chiu:2011qc,Becher:2011pf}, which differs from the one relevant for thrust in several important aspects. These differences arise because for jet broadening, the transverse momentum of the soft and jet radiation is of same size. 
In contrast, for thrust, or more general angularities with $a<1$, the transverse momentum of the soft radiation is parametrically smaller than that of the collinear radiation. For this reason soft recoil effects are power suppressed and the leading non-perturbative effects arise in the soft function. For broadening, on the other hand, there is a nontrivial interplay between the soft and jet functions and the leading non-perturbative effects affect both.

Since the characteristic scale of the soft and jet radiation is the same, the splitting among the two contributions is no longer unique for transverse observables such as jet broadening. One could introduce rapidity cutoffs to separate the different contributions, but it is more convenient to use analytic regulators. The soft and collinear functions then suffer from rapidity divergences which are not regularized by dimensional regularization. We have shown that these divergences only affect phase-space integrals and can be regularized using \cite{Becher:2011dz} 
\begin{equation}\label{regulator}
\int \!d^dk \,  \delta(k^2) \,\theta(k^0)\;\; \to\;\; \int \!d^dk \,  \delta(k^2) \,\theta(k^0) \,  \left(\frac{\nu}{k_+}\right)^{\alpha}\,,
\end{equation}
with $k_+ = k\cdot n$. Here, $n$ and its conjugate vector $\bar{n}$ are two light-cone reference vectors pointing in the direction of large momentum flow. The regulator $\alpha$ is introduced on the phase space of each particle in the final state. For the current analysis, it will be convenient to replace $\alpha\to \alpha/m$ for an $m$-particle final state in order to have homogeneous scaling for all phase-space integrals. The soft and jet functions then suffer from divergences in the analytic regulator which cancel in their product.

From consistency considerations \cite{Chiu:2007dg,Becher:2010tm}, one can show that the cancellation of the rapidity divergences induces a logarithmic dependence on the hard scale $Q$ in the low-energy matrix element, an effect called collinear anomaly \cite{Becher:2010tm}. More explicitly, the product of the soft and jet functions takes the form
\begin{equation}\label{collan}
\mathcal{J}(\mu)
\,\mathcal{J}(\mu)
\,\mathcal{S}(\mu) = \left( \frac{Q^2}{\mu^2} \right)^{-F(\mu)} 
W(\mu)\,,
\end{equation}
where the anomaly exponent $F(\mu)$ and remainder $W(\mu)$ are independent of $Q$. The $Q$-dependence is a pure power, and the result 
\eqref{collan} resums this dependence as long as the scale $\mu$ is chosen to be of the order of the transverse momentum. 

From the structure of \eqref{collan} one suspects that the dominant non-perturbative effects to transverse observables are corrections to the anomaly exponent $F(\mu)$. By performing an operator product expansion of the soft function and using consistency relations, we show that this logarithmic enhancement of non-perturbative effects is indeed present. For the case of jet broadening, we furthermore show that the leading non-perturbative corrections are determined by the same matrix element that governs the non-perturbative correction to thrust. 

\section{Jet broadening}

The soft function which occurs in the factorization theorem for broadening has the form \cite{Chiu:2011qc,Becher:2011pf}
\begin{align}\label{softfun}
 &{\cal S}(b_L,b_R,p_L^\perp,p_R^\perp) 
   = \sum\hspace{-0.65cm}\int\limits_{X, {\rm reg}}
   \delta(b_L- b_{X_L})\, \delta(b_R- b_{X_R} ) 	
   \\
   &\times \! \delta^{d-2}(p_L^\perp - \,p_{X_L}^\perp)\,
    \delta^{d-2}(p_R^\perp - \,p_{X_R}^\perp)
    \left| \langle X| S_n^\dagger(0)\,S_{\bar{n}}(0) |0\rangle \right|^2 \!\!,
    \nonumber
\end{align}
where the quantities
\begin{equation}
b_{X_{L/R}} = \frac12 \sum_{i\in X_{L/R}} |p_{i}^\perp|
\end{equation}
sum the transverse momenta of the particles in the left and right hemispheres with respect to the thrust axis. The subscript ``reg'' on the sum over intermediate states indicates that we have regularized this sum according to \eqref{regulator}. The soft emissions are described by two Wilson lines along the directions of the two jets. The soft function involves four $\delta$-functions. The first two set the values of the broadenings $b_L$ and $b_R$, and the latter two ensure that the transverse momentum of the soft radiation balances the one of the collinear radiation in each hemisphere.  

We now expand the soft function around the limit of large $b_L \sim b_R \sim |p_{L}^\perp| \sim |p_{R}^\perp| \gg \Lambda_{\rm QCD}$. Up to first order in the expansion, we obtain
\begin{align}\label{spert}
&{\cal S}(b_L, b_R ,p_{L}^\perp,p_{R}^\perp) 
   =\delta^{d-2}(p_{L}^\perp)\,\delta^{d-2}(p_{R}^\perp)  \\[0.2em]
   &\;\times \Big[  \delta(b_L) \delta(b_R)- \mathcal{M}_L \,\delta'(b_L) \delta(b_R)
   - \mathcal{M}_R \,\delta(b_L) \delta'(b_R) \Big]  \nonumber
\end{align}   
with
\begin{equation}
 \mathcal{M}_{L/R}
= \sum\hspace{-0.65cm}\int\limits_{X,{\rm reg}}\, b_{X_{L/R}} \left| \langle X | S_n^\dagger(0)\,S_{\bar{n}}(0) |0\rangle \right|^2\,,
\end{equation}
which corresponds to an expansion in moments of the soft function. The matrix element multiplying the first term in \eqref{spert} is trivial, since one sums over the final states without a constraint and the Wilson lines cancel by unitarity (after the analytic regulator is sent to zero). Note also, that the power corrections from the expansion in the transverse momenta $p_{L,R}^\perp$ must be second order because of rotation invariance in the transverse plane. 

To analyze the matrix elements $\mathcal{M}_{L/R}$, we follow Lee and Sterman \cite{Lee:2006nr} and rewrite them in terms of the transverse energy-flow operator,
\begin{align}
&\mathcal{M}_{L/R} = \frac12 \int d\eta \; \theta(\pm \eta) \\
 &\quad\sum\hspace{-0.65cm}\int\limits_{X,{\rm reg}} \langle 0 |\,S_{\bar{n}}^\dagger(0) S_n(0) \mathcal{E}_T(\eta) \,| X \rangle
    \langle X | S_n^\dagger(0)\,S_{\bar{n}}(0) |0\rangle\,,\nonumber
\end{align}
where we used the convention that left-moving particles with $k_-= k\cdot \bar{n}>k_+$ have positive rapidity $\eta$. The energy-flow operator $\mathcal{E}_T(\eta)$ is defined by its action on states $X$,
\begin{equation}
\mathcal{E}_T(\eta) \,| X \rangle = \sum_{i \in X} \left|p_{i}^\perp \right|
\delta(\eta-\eta_i) \,| X\rangle\,.
\end{equation}
We next perform a Lorentz boost along the thrust axis with rapidity $\eta'$. The Wilson lines are boost-invariant, but the presence of the analytic regulator in \eqref{regulator} spoils the invariance of the phase-space integrals. As a consequence, the sum over states picks up $m$ factors $e^{-\alpha \eta'/m}$, while the energy-flow operator transforms to $\mathcal{E}_T(\eta+\eta')$. By choosing $\eta'=-\eta$, the matrix element becomes independent of $\eta$ and the rapidity integration can be explicitly carried out. Expanding in the analytic regulator and performing the sum over final states, we end up with
\begin{equation}\label{expM}
\mathcal{M}_L = - \mathcal{M}_R = -\frac{\mathcal{A}}{2\alpha} 
+ \mathcal{O}(\alpha^0)\,,
\end{equation}
where
\begin{equation}
\mathcal{A}= \langle 0 |\,S_{\bar{n}}^\dagger(0) S_n(0) \mathcal{E}_T(0) S_n^\dagger(0)\,S_{\bar{n}}(0) |0\rangle
\end{equation}
is the same matrix element that drives the non-perturba\-tive shift of the thrust distribution, see \eqref{thrust:shift}. This is precisely the result which Lee and Sterman obtained when studying the angularities, but in our case $\alpha$ is a regulator, which is present in both the soft and the jet functions. Since QCD does not need the additional regulator to be well-defined, the pole in $\alpha$ must cancel against a divergence in the leading non-perturbative correction to the jet functions. 

\begin{figure}[t!]
\includegraphics[width=0.35\textwidth]{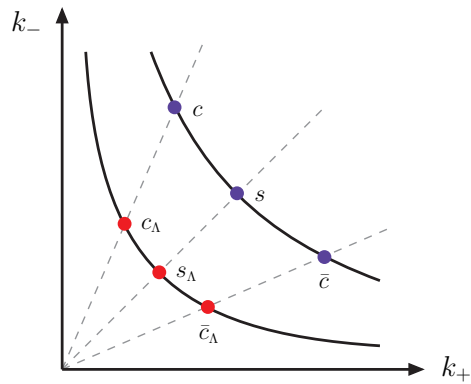}
\caption{\label{fig:modes}
Momentum modes for jet broadening. The virtuality of the perturbative modes on the upper hyperbola scales as $k^2\sim b^2$, whereas $k^2\sim \Lambda^2$ for the non-perturbative modes on the lower one. Note that perturbative and non-perturbative modes of the same type have the same rapidity, which implies the scalings $k_+^{c_\Lambda}\sim b\Lambda/Q$, $k_+^{s_\Lambda}\sim\Lambda$, $k_+^{\bar{c}_\Lambda}\sim Q\Lambda/b$.}
\end{figure}

To study this cancellation, we work in Laplace space,
\begin{equation}
\bar{f}(\tau_L, \tau_R) =  \int_0^\infty db_L db_R \,e^{-(\tau_L b_L+\tau_R b_R)} \,f(b_L, b_R)\,,
\nonumber
\end{equation}
where the cross section factors into a product
\begin{align}\label{Pdef}
    & \frac{d^2\sigma}{d\tau_Ld\tau_R}  \!= H(Q^2)\,\int_0^\infty\!\!\! dz_L dz_R
    \\[0.2em]
        &\quad\times\overline{\cal J}_{\!L}(\tau_L,z_L)\;\overline{\cal J}_{\!R}(\tau_R,z_R)\;
    \overline{\cal S}(\tau_L, \tau_R,z_L,z_R). \nonumber
\end{align}
 The variables $z_{L,R}\sim |x^\perp_{L,R}|$ arise from Fourier transforming the  transverse momentum variables $p_{L,R}^\perp$, see \cite{Becher:2011pf} for details.  It was shown in  \cite{Becher:2011pf} that the cancellation of the divergences in the analytic regulator $\alpha$ has to occur at the level of the integrand  before the $z_{L,R}$ integrations are carried out. Writing the soft function as a convolution of the perturbative result with a shape function describing the non-perturbative effects, the Laplace-space result reads
\begin{multline} \label{shapef}
 \overline{\cal S}(\tau_L,\tau_R,z_L,z_R) \\  
 =  \overline{\cal S}^{\rm pert}(\tau_L,\tau_R,z_L,z_R)
  \left[ 1+ \frac{\mathcal{A} }{2\alpha} 
 (\tau_L - \tau_R)  \right].
\end{multline}
An analogous representation holds for the jet functions 
\begin{equation}
 \overline{\cal J}_{\!L/R}(\tau,z) =  \overline{\cal J}^{\rm pert}_{\!L/R}(\tau,z) \left[ 1 -  \frac{\mathcal{B}_{L/R}}{2\alpha} 
 \,\tau \right].
 \end{equation}
Because the product of the jet and soft functions must be finite, we know that the non-perturbative part of the jet functions must have a $1/\alpha$ divergence which cancels the one encountered in the soft function: $\mathcal{B}_{L} = - \mathcal{B}_{R} = \mathcal{A}$ up to terms of $\mathcal{O}(\alpha^0)$. As the non-perturbative modes have different scalings in $k_+$ (see Fig.~\ref{fig:modes}), the cancellation of the rapidity divergences induces a logarithmic enhancement in the product of the soft and jet functions,
\begin{align}\label{scaling}
&  \frac{\mathcal{A} }{2\alpha} (\tau_L - \tau_R)  \left( \frac{\nu}{\Lambda} \right)^\alpha 
  -  \frac{\mathcal{A}}{2\alpha} \,\tau_L\left( \frac{\nu Q\tau_L}{\Lambda} \right)^\alpha
  \\[0.2em]
  &+ \frac{\mathcal{A}}{2\alpha} \,\tau_R \left( \frac{\nu}{Q\tau_R\Lambda} \right)^\alpha 
  = - \frac{\mathcal{A} }{2} \,
 \Big[\tau_L \ln(Q\tau_L) + \tau_R \ln(Q\tau_R)
 \Big]\,.\nonumber
\end{align}
Note that the rapidity divergences could not cancel if higher poles were present in the soft and jet functions. The single-pole structure is manifest in \eqref{expM}, but consistency requires it to be present for all anomalous observables. As a consequence, the leading power corrections are always enhanced by a single logarithm of $Q$. After the cancellation, the product $P = \overline{\cal J}_{\!L}\,\overline{\cal J}_{\!R}\,\overline{\cal S}$ becomes
\begin{align}\label{npAnomaly}
 &P(\tau_L, \tau_R,z_L,z_R)   \\
 &= P_{\rm pert}(\tau_L, \tau_R,z_L,z_R) \;
  (Q^2\tau_L^2)^{- \tau_L \mathcal{A}/4} \,
  (Q^2\tau_R^2)^{- \tau_R \mathcal{A}/4}\nonumber
  \end{align}
up to additional first-order power corrections which are not logarithmically enhanced. At higher orders, the cancellation of rapidity divergences implies that the logarithmic terms exponentiate, and the result \eqref{npAnomaly} takes a similar structure as the perturbative expression,
\begin{multline}
P_{\rm pert}(\tau_L, \tau_R,z_L,z_R) = W(\tau_L, \tau_R,z_L,z_R)\\
 \times (Q^2\tau_L^2)^{-F_{B}(\tau_L,z_L)}\, (Q^2\tau_R^2)^{-F_{B}(\tau_R,z_R)}
  \,,
\end{multline}
where $F_B$ is the anomaly exponent and $W$ the remainder function. We can thus view the enhanced power correction \eqref{npAnomaly} as a non-perturbative correction to $F_B$. 

To go back from Laplace to momentum space, we choose $\mu \sim 1/\tau_{L,R} \sim b_{L,R}$ and rewrite
\begin{equation}
(Q^2\tau_{L}^2)^{- \tau_L \mathcal{A}/4} = 1- \frac{\tau_L  \mathcal{A}}{2} \ln(Q\tau_L) = 1- \frac{\tau_L  \mathcal{A}}{2} \ln(Q/\mu)\,,
\nonumber
\end{equation}
up to power-suppressed terms which are not enhanced by logarithms of $Q$. Performing the inverse Laplace transform, we find that the power correction produces a shift of the distribution proportional to $\ln(Q/\mu) \sim \ln(Q/b_{L,R})$
\begin{align}
&\frac{d^2\sigma}{db_Ldb_R}(b_L,b_R) \\[0.2em]
&\quad=  \frac{d^2\sigma^{\rm pert}}{db_Ldb_R}\left(b_L- \frac{\mathcal{A}}{2} \ln \frac{Q}{b_L}, b_R- \frac{\mathcal{A}}{2} \ln \frac{Q}{b_R}\right)+ \dots
\nonumber
\end{align}
For the total and wide broadenings, $b_T = b_L + b_R$ and $b_W = \textrm{max}(b_L,b_R)$, it follows that the distributions get shifted by $\mathcal{A}\,\ln Q/b_T$ and $\mathcal{A}/2 \, \ln Q/b_W$, respectively. This is in agreement with the findings of \cite{Dokshitzer:1998qp}, in which the same effect was derived within the framework of the dispersive model. What our discussion adds is that it extends the QCD analysis of \cite{Lee:2006nr} to the case of the broadenings and shows that the result is model independent. What does not follow from our analysis, is that the same non-perturbative parameter ${\cal A}$ also governs the part of the power corrections which are not enhanced by logarithms of $Q$. Analyzing the non-logarithmic part would require an understanding of the structure of power corrections to the jet functions. We note that the universality established here holds for the definition \eqref{eq:rapdef}, but can be broken by hadron mass effects for alternative definitions \cite{Salam:2001bd,Mateu:2012nk}. This happens, for example, if the angularities are defined using pseudorapidity instead of rapidity.

\section{Other anomalous observables}

\begin{figure}
\centering
\begin{tabular}{lr}
\includegraphics[width=0.235\textwidth]{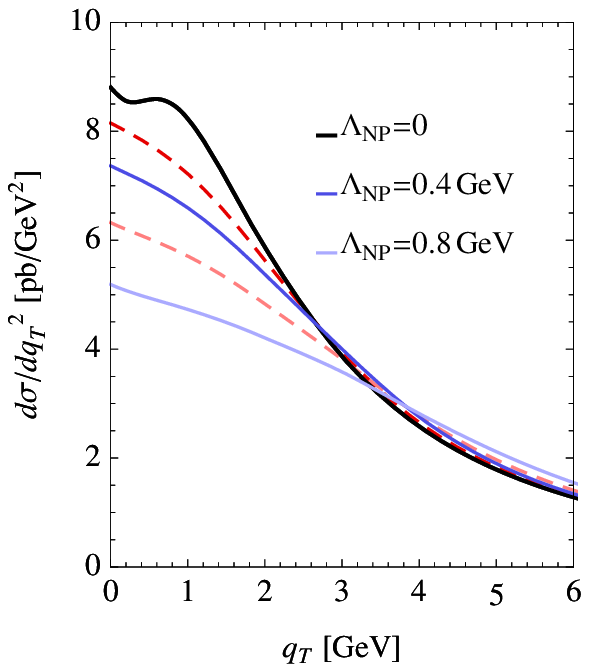}
& \includegraphics[width=0.235\textwidth]{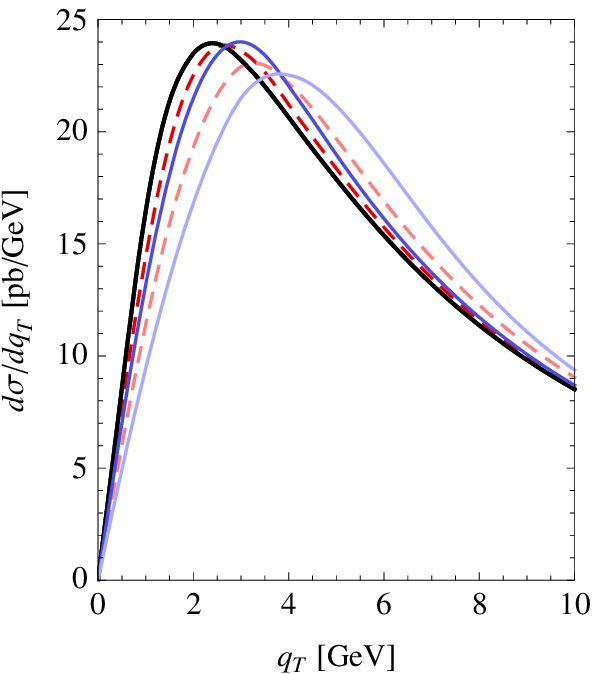}
\end{tabular}
\caption{Comparison of the non-perturbative corrections induced by the anomaly (solid blue lines) to the ones arising from a Gaussian cutoff  (dashed red lines). The plots show resummed predictions for $Z$ production at the Tevatron with the same input parameters and accuracy as in Figure 4 of \cite{Becher:2011xn}.
\label{qtNP}}
\end{figure}

The mechanism discussed above is at play in all observables which are affected by a collinear anomaly. Important examples include the spectra of electroweak bosons at low transverse momentum. The differential cross section for the scattering of hadrons $N_1$ and $N_2$ into a weak boson of mass $M$ with rapidity $\eta$ and transverse momentum $q_T^2=-q_\perp^2$ has the form \cite{Collins:1984kg,Becher:2010tm} 
\begin{multline}
\frac{d^2\sigma}{dq_T^2d\eta} =  \sum_{ij} H_{i j}(M^2)  \,\int d^2x_\perp \,e^{-i x_\perp q_\perp} \\
\left(\frac{x_T^2 M^2}{4 e^{-2\gamma_E}} \right)^{-F_{ij}(x_T^2)} B_{i/N_1} (\xi_1,x_T^2)\, B_{j/N_2} (\xi_2,x_T^2)\,,
\end{multline}
with $\xi_{1,2} =e^{\pm \eta}\, M/\sqrt{s}$. The sum runs over quark and anti-quark flavors for vector bosons, while $i=j=g$ for Higgs production. The $x_T$-dependence of the transverse parton distribution functions $B_{i/N} (\xi,x_T^2)$ as well as the anomaly exponent  $F_{ij}(x_T^2)$ can be computed perturbatively. Typically, non-perturbative effects are modelled by cutting off the integration over the transverse separation with a Gaussian, i.e.\ by replacing 
\begin{equation}
 B_{i/N} (\xi,x_T^2) \to e^{-\Lambda_{NP}^2 x_T^2} \,B_{i/N} (\xi,x_T^2)\,.
 \end{equation}
However, our analysis of jet broadening makes it clear that the leading non-perturbative effects are  due to a modification of the anomaly exponent
 \begin{equation}\label{powpt}
 F_{ij}(x_T^2) \to  F_{ij}(x_T^2) + \Lambda_{NP}^2 x_T^2\,,
  \end{equation}
where $\Lambda_{NP}^2$ can be extracted from the matrix element
\begin{equation}\label{eq:pTmat}
 \mathcal{M}_\perp
= \sum\hspace{-0.65cm}\int\limits_{X,{\rm reg}}\, p^2_{X_\perp} \left| \langle X | S_n^\dagger(0)\,S_{\bar{n}}(0) |0\rangle \right|^2\,,
\end{equation}
which implies that the correction only depends on the color representation, but not on the flavor, of $i$ and $j$~\footnote{With the regulator \eqref{regulator}, the matrix element \eqref{eq:pTmat} is scaleless since one can factor out an integral over total rapidity. The same would be true in the broadening case for $\tau_L=\tau_R$, see \eqref{shapef}. To work with a non-vanishing soft function one can use a symmetric form of the regulator in which the soft function is present, or split the radiation into hemispheres, see \cite{Becher:2013xia}.}.

The power correction \eqref{powpt} is of second order for the transverse momentum spectrum, while it is of first order for the broadening. However, other important aspects are the same for both observables. As explained after \eqref{scaling}, the relevant soft and collinear matrix elements can only have a single pole in the analytic regulator and the leading power correction is therefore enhanced only by a single logarithm of the hard scale $M$. This enhancement was predicted already in \cite{Collins:1984kg} on the basis of a model describing the emission of particles with low invariant mass at high rapidity. In this model, the scale in the logarithm is set by a non-perturbative parameter $\Lambda$ instead of the transverse separation $x_T \ll 1/\Lambda$  which governs the anomaly.

We  stress that the leading, logarithmically-enhanced terms exponentiate. This is important, since a strict twist expansion is worse than factorially divergent \cite{Becher:2011xn}. In Figure \ref{qtNP}, we compare the Gaussian cutoff to the enhanced non-perturbative effects from the collinear anomaly. We find that the effects are quite similar if the value of $\Lambda_{NP}$ is adjusted so that it will be difficult to distinguish the shapes with current data. However, in contrast to the ad-hoc models that are traditionally used, our analysis predicts the form of the power correction as well as its properties.

The mechanism described in this letter is universal and applies to all observables that are affected by a collinear anomaly. Our results provide a model-independent treatment of non-perturbative effects which is important for precision studies, both at lepton and hadron colliders.

{\em Acknowledgments:\/}
We are grateful to M.~Neubert, M.~Procura and J.~Thaler for discussions. T.B.\ and G.B.\ would like to thank the ESI Vienna for hospitality and support. The work of T.B.\ is supported by the Swiss National Science Foundation (SNF) under Grant No.\ 200020-140978. G.B.\ gratefully acknowledges the support of a University Research Fellowship by the Royal Society.

\end{document}